\documentclass[aps,showpacs,pre,amsmath,amssymb,floatfix,superscriptaddress]{revtex4}
\usepackage{amsmath}
\usepackage{amsfonts}   
\usepackage{graphicx} 

\pdfoutput=1

\begin{document} 
\title{Diffusion and Multiplication in Random Media}
\author{P. L. Krapivsky}
\affiliation{Department of Physics, Boston University, Boston, MA 02215, USA}
\affiliation{Institut de Physique Th\'eorique CEA, IPhT, F-91191 Gif-sur-Yvette, France}
\author{K. Mallick}
\affiliation{Institut de Physique Th\'eorique CEA, IPhT, F-91191 Gif-sur-Yvette, France}

\begin{abstract} 
We investigate the evolution of a population of non-interacting particles
which undergo diffusion and multiplication. Diffusion is assumed to be
homogeneous, while multiplication proceeds with different rates
reflecting the distribution of nutrients. We focus on the situation where  the distribution of
nutrients is a stationary quenched random variable, and show that the population exhibits
a super-exponential growth  whenever the nutrient distribution is unbounded. 
We elucidate a huge difference between the {\em average} and {\em typical} asymptotic 
growths  and emphasize the role played by the spatial correlations in the nutrient distribution.
\end{abstract}

\pacs{02.50.-r, 05.40.-a, 87.23.Cc}

\maketitle

\section{Introduction}
\label{main}

The evolution of a population in an inhomogeneous environment with spatially varying growth rates  
can display complex dynamical patterns, resulting from the competition between diffusion, random multiplication and possibly advection \cite{Zeldovich,ran_medium}. Inhomogeneities of the environment 
can greatly affect the population dynamics resulting in anomalous
spreading (neither diffusive, nor ballistic) and intermittent behaviors (for a review
on intermittency in random media  see, e.g.,  \cite{Zeldovichreview}) with  patches of population
growing in favorable places and  being  surrounded by large desertic 
regions.  Diffusion tails play the role of   vanguard agents that explore hostile  regions, 
in search of   ever-more hospitable locations to settle and to multiply.  
The offspring in the new settlement will eventually outgrow the original colony 
and generate a strong gradient current:
this can be seen   as a  migration of the whole  population.  A simple and  concrete  
example of this problem  is provided by bacteria  multiplying and diffusing 
in a petri dish where nutrients and inhibitors  are unevenly distributed, 
resulting in  complex  growth  patterns that are observable in actual 
experiments  \cite{Murray,Budrene}. In a  more abstract setting, the 
actual inhomogeneous space can be replaced by a rough phenotypic landscape in which the fitness
functional (that governs the growth  rate)  takes different values: here the coordinate 
variable  does not label a spatial location but rather the genetic content of each individual. 
This point of view has been inspired by S. Wright, R. A. Fisher and J. B. S. Haldane's classical  
studies in population genetic \cite{Ewens}. More recently it was developed by 
M. Eigen \cite{Eigen} as a  model for punctuated evolution of quasi-species 
and  has been further investigated by  W. Ebeling et al. \cite{Ebeling}, 
Y.-C. Zhang \cite{Zhang}, M. N. Rosenbluth \cite{R} and many other authors; 
see \cite{Baake,Krug} for a review of more recent work. 
 
In the present work, we examine a population of non-interacting particles that diffuse and
undergo the birth/death process depending on the availability of
nutrients. The population density $n({\bf x},t)$ evolves according to
the diffusion equation with a multiplicative noise:
\begin{equation}
\label{Diff-Lang} 
\frac{\partial n({\bf x},t)}{\partial t} = D\nabla^2
  n({\bf x},t) + \eta({\bf x})  n({\bf x},t)\qquad \text{with}\qquad n({\bf x},0) = n_0({\bf x}). 
\end{equation}

The diffusion coefficient $D$ is assumed to be uniform, while the birth/death rate 
$\eta({\bf x})$ is inhomogeneous. We shall focus on the simplest case 
of the stationary noise, $\eta=\eta({\bf x})$, and investigate how  spatial correlations 
of the noise affect the growth rate of  $n({\bf x},t)$.

The Langevin  equation \eqref{Diff-Lang} is linear, reflecting  the basic assumption 
that particles do not interact. The noise   $\eta({\bf x})$ is known through its stochastic 
properties and  changing these properties may drastically affect the behavior of  $n({\bf x},t)$.  
Equation \eqref{Diff-Lang} can be generalized in various ways, e.g. one can
take into account advection \cite{Nelson1,Nelson2,Nelson3,Schnerb}, 
study the evolution of a vector field in a random background (e.g. a magnetic field 
in the dynamo effect) \cite{ran_medium}, or consider several coupled fields as in the 
case of chemotaxis \cite{DeGennes,Verga}. 
One can introduce a non-linearity in order to take into account saturation 
effects \cite{ZeldPNAS}. Besides, it can also be interesting to consider situations with  
$\eta=\eta({\bf x},t)$, the dependence on time reflecting e.g. seasonal variations.

Although we shall chiefly employ a population dynamics vocabulary 
(particles,  nutrients, migrations), it is useful to keep in mind that the stochastic 
partial differential equation~\eqref{Diff-Lang} arises in many different contexts. 
In chemical physics, reaction kinetics is often modeled by equations similar 
to Eq.~\eqref{Diff-Lang} and these equations allow to predict macroscopic
patterns in the spatial distribution of reagents \cite{Mikhailov1,Mikhailov2}.
When the  amplification rate  $\eta(x)$ takes only negative values,
equation~\eqref{Diff-Lang} describes  non-interacting particles that diffuse
in a medium with random absorption. In the extreme case of 
$\eta(x) = - \sum_i \delta(x - x_i)$  with random positions $x_i$ of the traps, 
the trapping sites become perfect because 
$n(x,t)$ has to vanish for $x=x_i$. (More precisely, the above formulation 
applies in one dimension; in higher dimensions, the traps should be finite 
and  absorbing  conditions  are  set on the boundaries of the traps.) 
The literature on this classical subject is vast;  one of the most celebrated  
results asserts that the density decreases according to a stretched exponential law: 
$\ln n \sim - t^{d/(d+2)}$ in $d$ dimensions 
\cite{Varadhan,Balagurov,Grassberger,Redner,Anlauf,Theo,JMLTheo,JMLlivre}.

Another  important interpretation of  equation~\eqref{Diff-Lang} in terms
of polymer dynamics can be obtained  from  its  formal solution, 
\begin{equation}
\label{Green}  
n({\bf x},t) = \int d{\bf y} K({\bf x},t; {\bf y},0)\,  n_0({\bf y})\,,
\end{equation}
where the Green function $K({\bf x},t; {\bf y},0)$ is given by  the Feynman-Kac 
formula \cite{Schulman}
\begin{equation}
\label{Feynman}  
K({\bf x},t; {\bf y},0) = {\bf E_{y \to x}} \left(\!\exp\left[\int^{t}_{0}\eta({\bf x}(\tau)) d\tau\right] \right).
\end{equation}
The  expectation value ${\bf E_{y \to x}}$ with respect to the Wiener measure 
is taken over all  paths ${\bf x}(\tau)$ that begin at ${\bf y}$ at $t=0$ and end at ${\bf x}$ 
at time $t$. This formula can be rewritten in more familiar manner  as a path integral 
\begin{equation}
\label{pathintegral} 
  K({\bf x},t; {\bf y},0) = 
\int_{{\bf x}(0)={\bf y}}^{{\bf x}(t)={\bf x}}
\mathcal{D}{\bf x}(\tau)\, \exp\!\left\{\int^{t}_{0} d\tau \left[-\frac{\dot{\bf x}^{2}(\tau)}{4D}
+ \eta({\bf x}(\tau))\right]\right\}  .
\end{equation}
It is then natural to interpret  ${\bf x}(\tau)$ as a $d$-dimensional Gaussian  
polymer in the random potential $\eta({\bf x})$ \cite{Huse,Edwards,Cates,NR}, 
or as a  $d+1$-dimensional {\it directed} polymer with columnar
disorder \cite{KrugHH} (for a review on  directed polymers see \cite{HHZ}). 
The physics of directed polymers in  a random medium is closely related 
to the  growth of random surfaces. In one dimension, for instance, 
we write $n=e^m$ and recast \eqref{Diff-Lang} into $m_t = D(m_x^2 + m_{xx}) + \eta(x)$.  
Differentiating this equation with respect to $x$ and writing $h = - 2Dm_x$ we obtain
\begin{equation}
\label{BE}
 h_t + h h_x = D h_{xx} 
 + \xi(x)   \,\, \hbox{ where } \,\, \xi = -2D\eta_x \, .
\end{equation}
Thus after the transformation the noise becomes additive, but the governing 
Eq.~\eqref{BE} is now a {\em non-linear} stochastic partial differential equation. 
The Langevin equation~\eqref{BE} formally resembles the Kardar-Parisi-Zhang 
(KPZ)  equation \cite{HHZ,KPZ}, yet the noise $\xi$ in equation \eqref{BE} 
is  stationary, whereas in the KPZ equation the noise depends both on  
space {\it and}  time. [In the KPZ equation, the noise is typically taken to be 
the Gaussian white noise, $\langle\eta_{\rm KPZ}({\bf x},t) \eta_{\rm KPZ}({\bf x}',t')\rangle 
= 2\Gamma\,\delta({\bf x}-{\bf x}')\,\delta(t-t')$.] 
Hence the physical properties of the solutions to Eq.~\eqref{BE} are significantly 
different from those of the standard  KPZ equation \cite{Cates}. 

Equation~\eqref{Diff-Lang}  can also be viewed as a Schr\"odinger equation 
in imaginary time and with  a random potential;  it is  thus related  
to the physics of localization  \cite{JMLlivre,Lifshitz}.
In the present study $n(x,t)$ is a density of particles (not
a wave-function). For instance, in the no-noise case the total mass $\int dx\, n(x,t)$ 
is conserved, whereas  the analogous integral in the quantum case is not conserved. 
Furthermore, we are not directly  interested in eigenstates of~\eqref{Diff-Lang}  
but rather in the temporal behavior of its solution starting from a localized
initial state. All these differences explain why the phenomenology is rather different
from that of localization, although some of the  techniques developed to study
quantum disordered systems are useful in the analysis of Eq.~\eqref{Diff-Lang}.   

The behavior of solutions of  the apparently simple linear
Langevin equation \eqref{Diff-Lang} is not yet fully understood due to
a number of  puzzling features. One such feature is an astonishingly fast growth
(and sometimes a blow up that occurs in a finite time, or even
instantaneously \cite{R}). There are a few causes of these striking behaviors:
\begin{itemize}
\item The noise is multiplicative.
\item In many simple models, the  noise is unbounded.  Hence, the regions with  large
positive $\eta$  play a dominant role and lead to a  counter-intuitive
super-exponential growth. In this situation, the discretized (in space)
versions of Eq.~\eqref{Diff-Lang} differ  drastically   from the strictly
continuous version. 
\item The lack of self-averaging which is manifested by the huge
difference between {\em average} and {\em typical} behaviors.
\end{itemize}

The goal of this work is to analyze the Langevin equation~\eqref{Diff-Lang}  
when the noise is strongly correlated in space.  Some of our results are 
presented in Table~\ref{tablesynthese}  where we display  only  the scaling laws 
(numerical constants will be given throughout the text). The part of Table~\ref{tablesynthese}
describing the asymptotic growth laws for the uncorrelated Gaussian white noise
summarizes previous work. The notion of ``correlated'' noise is of course a bit vague. 
The one-dimensional case is exceptional as the very natural assumption 
that the {\em increments} of the noise are uncorrelated leads to the correlated
noise $\eta(x)$ which is essentially a trajectory of a random walk (in the lattice setting) 
or a Brownian motion (in a continuum setting), with $x$ playing a role to time. 
Table~\ref{tablesynthese} presents the asymptotic growth laws corresponding to such a noise. 
(Our main higher-dimensional results are collected in Table~\ref{tableHighDim}.)

\begin{table}  \centering
  \begin{tabular}{| c | c |}
  \hline
 Uncorrelated &   Correlated ($d=1$) 
\\ \hline                                                    &     \\ 
  {\bf Lattice  substrate}                          &   {\bf Random walk landscape}\\
  $e^{t^2}$ (Average,  all $d$)                &   $e^{t^3}$ \quad (Average)\\
  $e^{t {\sqrt{\ln t}}}$ (Typical,  all $d$)   &  $e^{t^{5/3}}$ ~~(Typical) \\
                          &     \\
 \hline                 &     \\ 
  {\bf Continuum  substrate}   &   {\bf Brownian landscape}\\
  $e^{t^3}$  ($d=1$)               &     $e^{t^5}$    \quad (Average)     \\
  Blow up $(d \ge 2)$             &  $e^{t^{5/3}}$  ~~(Typical)  \\
                             &     \\
 \hline           
\end{tabular}
\caption{Asymptotic growth laws for some of the population dynamics (numerical factors are omitted).  
The noise is assumed to be Gaussian, apart from the case when the noise forms a random walk 
landscape, that is, the increment of the noise are bounded. (The uncorrelated bounded noise is investigated  in Sect.~\ref{Bounded}.)
The results for the correlated noise in higher dimensions are discussed in Sect.~\ref{2d:corr}.}
\label{tablesynthese}
\end{table}

The rest of this paper is organized as follows. In Sect.~\ref{uncorrelated},
we review the behaviors when the noise is  either  uncorrelated or has short range correlations. We emphasize the emergence of puzzling behaviors and explain how 
seemingly contradictory results scattered in the literature can be synthesized in a coherent way.  In
Sect.~\ref{correlated} we analyze  the much  less studied situation of a correlated noise and  determine the growth law for the total population size in the one-dimensional setting. We then qualitatively describe  the situation in higher dimensions.

\section{Population dynamics with  short-range correlations}
\label{uncorrelated}

In this section we  assume that the local growth rate $\eta({\bf x})$ is on average
homogeneous. Therefore,  the mean value  $\mu = \langle\eta\rangle$ is constant
which can be set to zero; the general case is recovered 
by redefining  the local density:  $n({\bf x},t)\to n({\bf x},t)\, e^{\mu t}$. 
Further, because of  homogeneity, the spatial correlations of  $\eta$ must be translationally invariant.
The simplest assumption,  customary in studies of Langevin
equations,  is to consider  the random potential $\eta$ to be  uncorrelated
at different spatial locations. The fluctuating potential 
$\eta({\bf x})$ is thus taken to be a Gaussian white noise with zero average:
\begin{equation}
\label{noise} 
\langle\eta\rangle = 0, \qquad 
 \langle\eta({\bf x})
\eta({\bf x}')\rangle =  2\Gamma\delta({\bf x}-{\bf x}') \, . 
\end{equation}
The stochastic properties of the noise are now fully specified and equation~\eqref{Diff-Lang} 
defines  a well-posed problem which has been studied in numerous works 
\cite{Zeldovich,Zeldovichreview, Ebeling,Zhang,R,NR,GM,LW,wrong,T,Jaya} 
mostly in  one dimension. The conclusions of these studies, established 
through various  methods and approximations, seemed initially contradictory. 
However, the issue was settled in  \cite{R} by an exact analytical calculation based 
on a minimax variational principle. In the large time limit,  the dominant contributions 
to the population density arise  from very small, rapidly growing isolated regions. Although
these regions are very rare and highly improbable, they produce high density peaks that 
dominate the whole statistics. This has led to the following unexpected asymptotic 
behavior in one dimension: 
\begin{equation}
\label{asymp} 
\ln \langle n(0,t)\rangle \to  \frac{\Gamma^2 t^3}{12 D}\,\,.
\end{equation}
This super-exponential growth has been also found 
by estimating Brownian motion expectations \cite{LW}  and by  a path-integral approach 
that has additionally allowed to calculate the pre-factors of the exponential behavior \cite{T}:
\begin{equation}
  \langle n(0,t)\rangle \to 
\frac{ \Gamma^2 t^{5/2}} {2 (\pi D^3)^{1/2}}\,
 \exp\!\left( \frac{\Gamma^2 t^3}{12 D} \right).
 \label{result:RTao}
\end{equation}
 
In higher dimensions, the behavior is even more puzzling. It has been argued 
\cite{R,EPGross} that in  two  dimensions, a  divergence occurs at a finite time 
$t_c\sim D/\Gamma$; for $t > t_c$ the solution blows up.  Further, when $d>2$, 
the divergence is instantaneous, i.e., equation~\eqref{Diff-Lang} with Gaussian 
white noise \eqref{noise} is ill-defined. These striking behaviors exhibited by 
the Langevin equation \eqref{Diff-Lang} with Gaussian noise \eqref{noise} are 
consequences of two major properties of the noise: The Gaussian noise
is both uncorrelated and unbounded.  We now  investigate the consequences
of relaxing these assumptions.

\subsection{Taming the White Noise: Lattice Regularization or 
Finite Short-Range  Correlations}

The white noise displays totally uncorrelated  fluctuations 
at all scales. However,  in the present context of a population dynamics
it is natural to assume some spatial  coherence  in the  local
varying  conditions and that the noise has a non-vanishing  correlation length. 
The simplest manner to implement this model  is to put the
system on a lattice and to assume that  the noise is uncorrelated at  different
lattice sites. The  effective correlation length is thus  equal to
the lattice spacing. Mathematically, one has to solve the discrete set of equations 
\begin{equation}
\label{nj-eq} 
\dot n_{\bf j} = D\nabla^2 n_{\bf j} + n_{\bf j}
\eta_{\bf j}
\end{equation} where ${\bf j} = (j_1,\ldots,j_d)\in \mathbb{Z}^d$ if
the lattice is hyper-cubic.  Further, the operator $\nabla^2$ denotes
the discrete Laplacian (e.g.  in one dimension we have  $\nabla^2 n_j = n_{j-1}-2n_j+n_{j+1}$).
 The noise in \eqref{nj-eq} is Gaussian with
 the following characteristics: 
\begin{equation}
 \langle\eta\rangle = 0, \qquad  \langle\eta_{\bf i}
\eta_{\bf j} \rangle =  2\Gamma\delta_{{\bf i}, {\bf j}} \, .
\end{equation}
In this case, the growth is chiefly universal, it is independent of the spatial dimension (up to numerical constants) and is given by $\langle n \rangle \sim e^{\Gamma t^2}$  \cite{Zeldovich}. 
Hereinafter, we shall use the convention that in the asymptotic such as $n\sim \exp(At^a)$, 
the displayed term gives the correct {\em controlling} exponential factor, so that the actual 
asymptotic may be something like $n\sim Bt^b \exp(At^a)$. Thus the asymptotic 
$\langle n \rangle \sim e^{\Gamma t^2}$ is actually a shorthand formulation of the leading asymptotic
of the logarithm:
\begin{equation}
\label{lattice}
\lim_{t\to\infty}\frac{\ln  \langle n \rangle}{\Gamma t^2} = 1\,.
\end{equation}

The growth law \eqref{lattice} involves averaging over the
disorder \eqref{averaging}. The simplest set-up that automatically
enforces such an averaging occurs in the situation when the evolution
begins from the uniform initial condition: $n_{\bf j}(0)=1$ for all
${\bf j} \in \mathbb{Z}^d$. Indeed, no averaging is needed because,  for the
infinite lattice,  all values of the noise are appropriately
sampled. Thus,  we can ignore diffusion altogether \cite{Zeldovich}. Then
$n=e^{\eta t}$, so that
\begin{equation}
\label{averaging} 
\langle n \rangle = \langle e^{\eta t} \rangle =
\int_{-\infty}^\infty \frac{d\eta}{\sqrt{4\pi\Gamma}}\,
\exp\!\left(\eta t - \frac{\eta^2}{4\Gamma}\right)  = e^{\Gamma t^2} \,. 
\end{equation} 

If, however, the initial condition is  localized, e.g. $n_{\bf j}(0)=\delta_{{\bf j}, {\bf 0}}$, the behavior
\eqref{lattice} will arise only after averaging over all
distributions of the disorder. The typical behavior with a fixed noise, however, 
differs  greatly: this is the sign of the lack of self-averaging. To establish the typical growth law, we 
denote by $L$ the  size of
the domain visited by the particles. This domain contains about $L^d$
sites. The largest noise $\eta_*(L)$ at  these sites is evaluated using
the extreme statistics criterion (see e.g. \cite{book}) to give
\begin{equation} 
\int_{\eta_*}^\infty
\frac{d\eta}{\sqrt{4\pi\Gamma}}\, \exp\!\left( -
\frac{\eta^2}{4\Gamma}\right)\sim \frac{1}{L^d} \, , 
\label{eq:criterion}
\end{equation}
 from which
\begin{equation} 
\eta_*(L) \simeq \sqrt{4 d \,  \Gamma \ln L} \, .
 \label{eq:typicaletaL}
\end{equation}
We now ought to find out how the size  $L$ of the domain visited by a 
particle grows with time. A naive  estimate postulates a diffusive scaling law 
$L \sim \sqrt{t}$, thereby leading to 
$n_{\text{typ}} \sim \exp( \eta_*(L) t )\sim  \exp\!\big( t \sqrt{2 d \,  \Gamma \ln t}\big)$. 
This is wrong, however,  as explained e.g. in Ref.~\cite{Zeldovich}.
The correct  argument  proceeds  by  averaging the optimal growth  
$\exp( \eta_*(L) t )$ for a given path of  lengths $L \ge 0$ over all possible 
paths  weighed  by their probability of occurrence. 
Therefore, the  typical population size grows as
\begin{equation}
 n_{\text{typ}} (t) \sim  \int_0^\infty \frac{dL}{(4\pi
Dt)^{d/2}}\, \exp\!\left(t \sqrt{4 d \,\Gamma \ln L} - \frac{L^2}{4Dt}\right)
 \, .
\label{Saddle1}
\end{equation}
We  calculate this integral by the saddle-point method. The exponent has 
a sharp maximum at
\begin{equation} 
 L\simeq \frac{T}{(\ln T)^{1/4}}\,, \qquad T\equiv 
 t D^{1/2}(4d\Gamma)^{1/4} \, . 
 \label{OptimalL1}
\end{equation}
Keeping only the dominant exponential factor in \eqref{Saddle1} we obtain 
\begin{equation}
\label{typical}
  n_{\text{typ}} (t) \sim   e^{t \sqrt{4 d \, \Gamma \ln T}} \, . 
\end{equation}
This  typical growth is essentially universal (the spatial dimensionality appears only
in amplitudes) and the growth is just barely faster than
exponential. It is important to note  that  {\it the typical population growth rate}  
results from   particles that follow {\it optimal paths} that are almost ballistic, 
as seen from equation~\eqref{OptimalL1}, rather than diffusive; 
these optimal paths are therefore {\it highly non-typical individual trajectories}. 
Here, the naive estimate  $L \sim \sqrt{t}$ does, by chance,   provide 
the correct  functional dependence  $ t \sqrt{ \ln t} $ 
inside the exponential in equation~\eqref{typical},  but 
with a  coefficient wrong by a factor $\sqrt{2}$. In the next section, we shall encounter
cases where the naive estimate leads to erroneous results. 

The average population  growth, given in equations~\eqref{lattice}  or~\eqref{averaging},
radically  differs  from the typical growth~\eqref{typical}. 
However, these two results can be reconciled  as follows.
We have established~\eqref{typical} by estimating the
value of  $\eta_*(L)$, the largest noise that occurs amongst  $L$
sites, using  the criterion~\eqref{eq:criterion}. Yet the value  given in~\eqref{eq:typicaletaL}  
for  $\eta_*(L)$  is valid for a {\it  typical realization of the noise $\eta(x)$}.
In fact, the  average  growth of the population, given by  $\langle n \rangle,$   
is dominated by {\it highly non-typical realizations of the noise} that
must be taken into account: in order to  calculate $\langle n \rangle$ correctly 
we have to let {\em both} the path and the background noise
fluctuate.  For $L$ independent realizations of the
Gaussian random variable $\eta$, the cumulative  distribution
of the maximum  $M$ is given by
\begin{equation}
 \rm{Prob}(\eta_{max}  \le M) = \left( \,\,\,  \int_{-\infty}^{M} 
 \frac{d\eta}{\sqrt{4\pi\Gamma}}\, {\rm e}^{ -
\eta^2/{4\Gamma} }\,\,\, \right)^L   \, .
\end{equation}
Taking the derivative of this expression, we find that the probability distribution of $M$ 
 \begin{equation}
  P_L(M)  =  L  
 \frac{{\rm e}^{ -M^2/{4\Gamma} } }{\sqrt{4\pi\Gamma}}\,
 \left( \,\,\,  \int_{-\infty}^{M} 
 \frac{d\eta}{\sqrt{4\pi\Gamma}}\, {\rm e}^{ -
\eta^2/{4\Gamma} }\,\,\, \right)^{L-1} \sim  \,
  {\rm e}^{ -M^2/{4\Gamma} } \, ,
\end{equation}
where the last expression, in which we have retained  only the controlling 
exponential factor, is valid for large values of $M$.  Since
the average population at time $t$  over a range of $L$ sites grows as ${\rm e}^{M t}$, 
we obtain 
\begin{equation}
   \langle n \rangle \sim  \int_0^\infty dL \,
  \frac{ \exp\!\left( - \frac{L^2}{4Dt}\right) }{(4\pi Dt)^{d/2}}\,
  \int   dM \,{\rm e}^{M t}   \, {\rm e}^{ -M^2/{4\Gamma} }   \, .
\end{equation}
The asymptotic is evaluated using the saddle-point technique to yield
\begin{equation}
  \langle n \rangle   \sim {\rm e}^{\Gamma t^2}  \, , 
\end{equation}
in agreement  with equation~\eqref{averaging} which was obtained in the zero-dimensional case.
We note that the diffusion constant $D$ appears neither in the average behavior nor in the typical behavior~\eqref{typical}:  it affects only the sub-leading corrections.

Thus the short-ranged correlations drastically modify the  behavior of the solutions 
to Langevin equation \eqref{Diff-Lang}  by regularizing the noise term. 
The short-range fluctuations that were responsible for the blow-up
in dimensions $d \ge 2$ are suppressed and, on the lattice,  
equation~\eqref{Diff-Lang} is  well defined in {\it all dimensions}.  The growth of $n(x,t)$
is the  universal Gaussian  law~\eqref{lattice} that does not change with dimension.
We finally note that another way to regularize  equation~\eqref{Diff-Lang} without
discretizing space is to consider a colored Gaussian noise instead
of a white noise. A frequently used  example is the Gaussian  Ornstein-Uhlenbeck  noise
with correlations given by $\langle \eta(x) \eta(x') \rangle = \frac{\Gamma}{\xi}\,\exp(-|x -x'|/\xi)$.
This noise  has exponentially decaying correlations with correlation length $\xi.$   
For such a  noise, the population grows as $\langle n  \rangle \sim e^{\Gamma t^2/2\xi}$ 
(see \cite{LW}). This asymptotic is again independent of  the diffusion constant and the 
dimensionality of space.

\subsection{Taming the White Noise: Bounded Noise Distributions}
\label{Bounded}

The lack of the upper bound for the Gaussian noise is an obvious reason for the
appearance of the faster-than-exponential growth found in  \eqref{asymp},
\eqref{lattice}, and \eqref{typical}. For a bounded  noise with 
$\eta\leq\eta_{\text{max}}$, the growth cannot be faster than $e^{\eta_{\text{max}} t}$.  
Interestingly, in most cases the controlling factor is equal to  $e^{\eta_{\text{max}} t}$ 
and  the spatial dimensionality   or   details of the noise distribution 
(such as the behavior of the noise distribution 
function $\rho(\eta)$ in the proximity of $\eta\leq\eta_{\text{max}}$)
play a secondary role, namely they affect the pre-factor in the growth law. 
Let us look at this pre-factor. To appreciate its behavior, it suffices to analyze 
noise distributions with a finite number of different values of the noise. Without loss 
of generality we set the maximal noise to unity and write
\begin{equation}
\label{noise:discrete} 
\rho(\eta) = p\delta(\eta-1) + \sum_{i=1}^n
q_i\delta(\eta-a_i)
\end{equation} with
\begin{equation} p+\sum_{i=1}^n q_i=1, \quad a_1,\ldots,a_n<1 \, .
\end{equation}
If the diffusion and the  multiplication process occur  on a
one-dimensional lattice, the lattice can be thought to be  an array of
domains where the noise is maximal. Adjacent domains are separated by
sites where the noise is smaller. The probability density
of domains of length $k$ is given by
\begin{equation}
\label{segment} 
\Pi_k = (1-p)^2 p^k
\end{equation} 
where the factor $p^k$ accounts for $k$ consecutive
sites with maximal noise and the factor $(1-p)^2$ assures that the
noise at  the boundary sites is smaller than 1. (Using \eqref{segment}
one can compute the fraction of lattice with maximal noise to yield
$\sum_{k\geq 1}k\Pi_k=p$ as it should be.)

Consider the simplest situation where  the entire lattice is  initially
uniformly filled: $n_j(0)=1$ for all $j$. The asymptotic behavior
can be quantified by the average density
\begin{equation}
\label{n-av} n(t) = \lim_{L\to \infty} \frac{1}{L}\,\sum_{j=1}^L n_j(t) \, . 
\end{equation}
We first observe that the average density has the trivial upper bound
\begin{equation}
\label{upper} n\leq e^t \, .
\end{equation}
We now  construct  a lower bound for the average
density. The idea is to consider the evolution on domains where the
noise is maximal and use the absorbing boundary conditions on the ends
of each domain. This is an obvious lower bound, yet we will see that
it exhibits largely the same growth as the   upper bound. 
To proceed,  we make the  assumption
(to  be confirmed a posteriori) that the chief asymptotic is
actually provided by very long domains ($k\gg 1$). For such domains we
can replace the discrete diffusion equation by a  continuous one and 
we need to solve
\begin{equation}
\label{Diff} 
\frac{\partial n}{\partial t} = D\nabla^2 n + n
\end{equation} 
on the interval $0<x<k$ subject to the initial
condition $n(x,0)=1$ and the absorbing boundary conditions
$n(0,t)=n(k,t)=0$ for $t>0$. In the long-time limit, the spatial
distribution approaches the smallest eigenfunction of the
Laplace-Dirichlet operator: $n=f(t)\sin(\pi x/k)$. Plugging this into
\eqref{Diff} we get  $n \sim e^{t(1-D\pi^2/k^2)}\, \sin\tfrac{\pi x}{k}$. 
Using Eqs.~\eqref{segment} and  \eqref{n-av},
we arrive at the estimate for the lower bound:
\begin{equation}
\label{sum} 
n \sim \sum_{k\geq 1}\Pi_k k\,e^{t(1-D\pi^2/k^2)}
  \sim  \int_0^\infty dk\,k\,e^{t(1-D\pi^2/k^2)-k\ln(1/p)} \, ,
\end{equation}
where we have replaced the sum by an integral because the
asymptotic is dominated by the contribution of large domains. 
This integral  can be calculated by the saddle-point method.
One finds that the exponential term in the
integrand has a sharp maximum at $ k_*=\left[2\pi^2Dt/\ln(1/p)\right]^{1/3} .$
Keeping only the leading and sub-leading terms,  we arrive at the lower bound
\begin{equation}
\label{lower} 
n >  \exp\!\left\{t-\frac{3}{2}\,(2\pi^2Dt)^{1/3}[\ln(1/p)]^{1/3}\right\} \, .
\end{equation} 
Comparing the upper and lower bounds,
Eqs.~\eqref{upper} and \eqref{lower}, we see that the controlling
exponential factors are the same. This provides an evidence in favor
of the general assertion that for an arbitrary bounded noise in
arbitrary dimension the controlling exponential factor is universal
and determined by the maximal noise:
\begin{equation}
\label{bound_noise}
 n\sim e^{\eta_{\text{max}} t}\,.
\end{equation}

The  above derivation of the lower bound \eqref{lower} can be generalized to an arbitrary
dimension. We again consider domains of neighboring sites with maximal
noise. In principle, there can be an infinite domain (when the density
$p$ of the maximal noise exceeds a percolation threshold
$p_c(d)$). Let $p<p_c(d)$: If in this situation the controlling
exponential factor is still given by \eqref{bound_noise}, it will
certainly be valid for larger $p$. When $p<p_c(d)$, the domains are
finite and generally small. We now proceed as before, namely we set
$n=0$ outside the domains as this will obviously provide a lower
bound. A well-known argument (see \cite{book}) implies that the
domains which lead to the largest contribution are balls, so one must solve
\begin{equation}
\label{Diff-d} 
\frac{\partial n}{\partial t} =
D\left(\frac{\partial^2 n}{\partial r^2}+\frac{d-1}{r}\,\frac{\partial
n}{\partial r}\right) + n  \, , 
\end{equation} 
inside the ball $r\leq R$ with the absorbing boundary
condition $n(r=R,t)=0$ on its surface. The solution reads
$ n\sim e^{t(1-\lambda_1^2 D/R^2)}\,J_\delta(\lambda_1 \tfrac{r}{R}) \,, $
where $J_\delta$ is the Bessel function with index
$\delta=(2-d)/2$ and $\lambda_1$ is the first zero of this Bessel
function. Proceeding as in one dimension one arrives at a lower bound
\begin{equation} 
n> e^t \int_0^\infty
dR\,R^{d-1}\exp\!\left\{-\lambda_1^2
\frac{Dt}{R^2}-\ln(1/p)V_dR^d\right\}
\end{equation} 
where we have taken into account the fact that the probability
that all sites of a ball of radius $R$ have maximal noise scales as
$p^{V_dR^d}$ for large $R$ (here $V_d$ is the volume of a unit
ball). Estimating the integral we obtain
\begin{equation} 
n> e^t\,\exp\!\left\{-C_d (Dt)^{d/(d+2)}\right\} 
 \hbox{  with }
 C_d=\frac{d+2}{d}\left(\frac{d\lambda_1^d V_d
\,\ln(1/p)]}{2}\right)^{2/(d+2)} \, .
\end{equation}

For the continuous noise distributions, the behavior near the 
maximal noise plays a  crucial role.
We have analyzed noise distributions that behave as
\begin{equation} 
\text{Prob}(1-\epsilon<\eta<1)=A \epsilon^a, \quad a>1
\end{equation} 
in the $\epsilon\to 0$ limit. The corrections to the controlling exponential factor 
\eqref{bound_noise} are  similar to the case of the discrete noise distributions, 
e.g. in one dimension
\begin{equation}
n>e^t\, \exp\!\left\{-C(a,A) t^{1/3} (\ln t)^{2/3}\right\}. 
\end{equation}

\section{Population Dynamics in a Brownian Landscape}
\label{correlated}

In this section, we relax the  unrealistic assumption that the noise is uncorrelated when the 
distance exceeds a certain threshold. Within the ecological interpretation where the noise
refers to local conditions, it is natural to assume that conditions
change from site to site, yet if somewhere conditions are very good,
they are also very good in the proximity. This suggests to consider a model where $\eta(x)$ is 
a random landscape. In one dimension, one practical realization of such a  landscape 
is to  take $\eta(x)$ to be  a random walk  trajectory (in the lattice setting) or
a Brownian trajectory (in a continuum setting). In higher dimensions, $\eta$ will be taken to be a Gaussian field. In all cases, the roughness of the surface defined by the noise governs
the population  growth law.

\subsection{One-Dimensional Case: Heuristic Analysis}

In one dimension, $\eta(x)$  is taken to be a Brownian curve, in which
 $x$ plays the role of a `time' variable.  We  assume that the initial population seed is located at the origin
$n(x,t=0)=\delta(x)$ and we  set $\eta(x=0)=0$. Then the noise is given by 
\begin{equation} 
 \eta(x) = \int_0^x \xi(u) du
\end{equation}
where $ \xi $ is a Gaussian white noise. Thus, the  autocorrelation of the landscape reads
 \begin{equation}
 \langle \eta(x)^2 \rangle =  2\Gamma |x| \,.
\end{equation}
In the following we normalize the Brownian  landscape by replacing  $\eta(x)$ 
by $\sqrt{2\Gamma}\eta(x)$. 

First, let us estimate the typical and the average growth laws by employing a heuristic 
reasoning. Heuristic arguments elucidate the physical mechanisms leading to 
the super-exponential growth and shed light on the crucial distinction between 
the average and typical behaviors. 

To estimate the typical growth, one could argue that the particle
visits roughly $L  \sim \sqrt{t}$ different lattice sites (in one dimension)
during the time interval $(0,t)$ and that  highest value of the  Brownian noise  among
these sites is $\eta_{\rm max}\sim \sqrt{L} \sim
t^{1/4}$. Because the  density must grow as $e^{\eta_{\rm max} t}$, one
could anticipate that $n \sim e^{t^{5/4}} \, .$
However, the  above derivation is too rough  even for a heuristic
argument: as already discussed in the previous section, 
the Feynman-Kac formula~\eqref{pathintegral}, representing  a formal solution to
equation~\eqref{Diff-Lang}, contains  a summation over all possible trajectories. 
Therefore we should let $L$ fluctuate. Thus we write
\begin{equation}
\label{typ_estimate}  
n_{\rm typ}\sim \exp\!\left[t\sqrt{L}
-\frac{L^2}{4t}\right] 
\end{equation}
and maximize with respect to  $L$ to give $L\sim t^{4/3}$ leading to
\begin{equation}
\label{density}
  n_{\rm typ} \sim e^{t^{5/3}} \, .
\end{equation}
Interestingly, the optimal length  $L \sim t^{4/3}$ is super-ballistic.  

Another argument  leading to the  same result proceeds by  saying  that,
since the total number of particles grows very
rapidly, the total number of visited sites actually grows faster than
diffusively.  If all particles which are present in  the system at
time $t$ were initially at the origin (an admittedly rough
assumption) we estimate the size $L$ of the segment of sites visited
by the particles from the criterion $\frac{n}{\sqrt{4\pi t}}\, e^{-L^2/(4t)}\sim 1$.
This gives (we take into account that $n$ grows 
exponentially, i.e., much faster than a  power law) $ L\sim \sqrt{t\,\ln n} \, .$  
The maximal noise on this segment is $\eta_{\rm max}\sim \sqrt{L}\sim (t\,\ln n)^{1/4}$. 
Now using $n\sim e^{\eta_{\rm max} t}$ we obtain 
$ \eta_{\rm max}\sim t^{-1}\,\ln n\,.$
Combining these two expressions for $\eta_{\rm max}$, we get
$t^{1/4}(\ln n)^{1/4}\sim t^{-1}\,\ln n$, leading to \eqref{density}.

The typical growth law \eqref{density} is obtained for a given realization of 
the potential $\eta(x)$. The average growth of the population,  averaged over different  
realizations of the potential, is very different, namely it is much faster.
Again, the very rare fluctuations of the landscape dominate the average. 
The simplest way to estimate the  average 
growth is to keep two free parameters, the size $L$ of the segment
visited by the particles and the maximum $M$ reached by the noise
$\eta(x)$ on this segment. This leads to 
\begin{equation}
\label{av_est}  
\langle n\rangle\sim \exp\!\left[t M - \frac{M^2}{2 L}
-\frac{L^2}{4t}\right] \, , 
\end{equation}
where the factor $\exp(-M^2/2L)$ represents
the  tail of the distribution of the maximum of a Brownian path over a range $L$.
The  maximum   distribution of a Brownian path is a classical result
that can be derived via the image method \cite{BM}.
Maximizing in $L$ and $M$ we get $L\sim t^3,  M\sim
t^4$ and then Eq.~\eqref{av_est} results in
\begin{equation}
 \langle n\rangle\sim e^{t^5} \, .
\label{av_cont} 
\end{equation}

The asymptotic growth laws \eqref{density} and \eqref{av_cont} crucially depend on 
the assumption that the noise is the Brownian landscape. To illustrate other possible
behaviors we give two examples. 

\subsubsection{Random walk landscape}

In this situation, the population dynamics occurs on the one-dimensional lattice and 
$\eta$ is assumed to be a random walk rather than a Brownian motion: 
$\eta_{j+1} - \eta_j = \pm 1$, where $\pm$ are chosen independently and with equal
probabilities so that the surface $\eta(x)$ has no tilt. The
noise can go up $L$ steps in a row (albeit with a small probability
$2^{-L}$) and this rare fluctuation provides the dominant
contribution. Indeed, writing
\begin{equation}
\label{av_estimate}  
\langle n\rangle_{\rm rw}\sim \exp\!\left[t
L -\frac{L^2}{4t} - L\ln 2\right] \, , 
\end{equation} 
we see that in the $t\to\infty$ limit the exponentially
small probability of the rare fluctuation is totally outweighed by its
huge contribution. Maximizing in $L$ we obtain $L\sim t^2$ and then
Eq.~\eqref{av_estimate} leads to
\begin{equation}
\label{av_lattice} 
  \langle n\rangle_{\rm rw}\sim e^{t^3} \, .
\end{equation}
The difference between \eqref{av_cont} and \eqref{av_lattice} is the consequence of the fact that the noise increments are {\em bounded} for the random walk landscape.

\subsubsection{Fractional Brownian motion}  

Consider the case of a self-affine disordered landscape, in which Gaussian fluctuations grow 
with distance with a positive Hurst exponent $0<H<1$, 
namely $\langle \eta(x)\eta(y) \rangle = |x - y |^{2H}$. (The localization of a
quantum particle in such self-affine potentials has been recently studied, see  
\cite{Brazil,Russ,JML} and references therein.)  Then $\eta_{\rm max}\sim L^H$ 
and the same reasoning as above leads to the growth law 
\begin{equation}
\ln n_{\text{typ}}\sim t^{(2+H)/(2-H)}\,.
\end{equation}

\subsection{One-Dimensional Case: Quantitative Analysis using the WKB Method}

In this subsection,  we derive the asymptotics \eqref{density} and \eqref{av_cont}. 
Equation \eqref{Diff-Lang} is linear, so it is useful to perform a spectral decomposition.  
We write
\begin{equation}
 n(x,t) = \int dE \, {\rm e}^{Et}\, C(E)\, n_E(x) 
\label{SpectAnalysis}
\end{equation}
where the eigenfunction $n_E(x)$  satisfies a Schr\"odinger equation
\begin{equation}
\label{eigenEq} 
 E n_E(x) = D\frac{d^2  n_E(x)}{d x^2} 
   + \sqrt{2 \Gamma} \eta(x) n_E(x) \, , 
\end{equation}
with normalized Brownian landscape $\eta(x)$ playing the role of a potential. 
We also impose the normalization condition on the total mass of the eigenfunction 
\begin{equation}
\int n_E(x) dx = 1 \, .
\label{norm}
\end{equation}
The coefficient  $C(E)$ in equation \eqref{SpectAnalysis} is determined
by the initial condition. Using $n(x,0) = \delta(x)$
and the fact that the eigenfunctions $ n_E$ are mutually orthogonal,
we obtain  $C(E) = n_E(0)$  and this allows us to write
\begin{equation}
 n(x,t) = \int dE \, {\rm e}^{Et}  n_E(0) n_E(x) \, . 
\label{SpectAnalysis2}
\end{equation}
Integrating over $x$ and using the normalization~\eqref{norm} yields
\begin{equation}
 n(t) = \int dE \, {\rm e}^{Et}  n_E(0)  \, .
\label{PopTotale}
\end{equation}

\begin{figure}[ht]
\begin{center}
\includegraphics[width=0.45\textwidth,angle=-90]{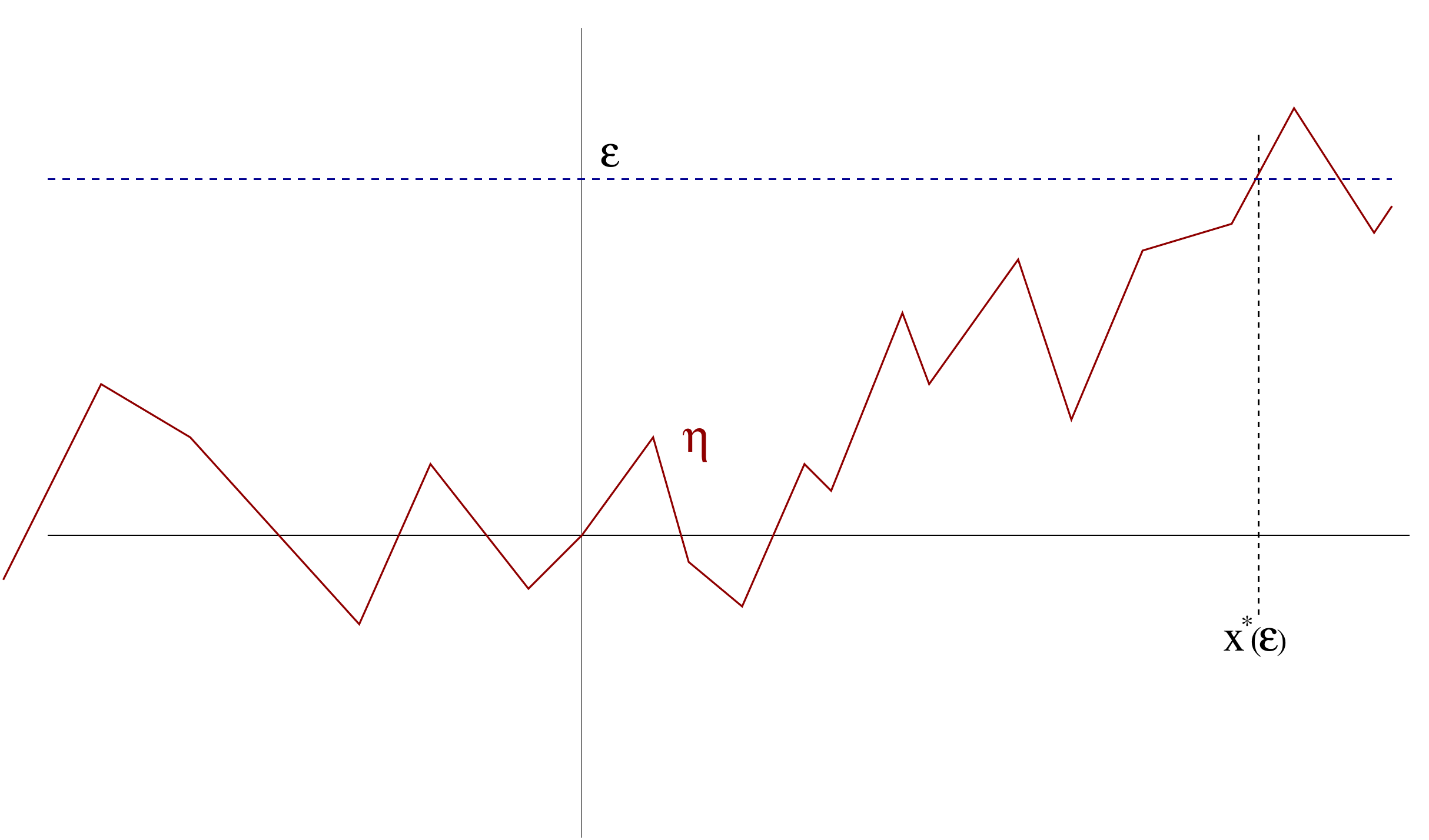}
\caption{An illustration of the WKB method: the random landscape $\eta(x)$ starting
 from the value 0 at the origin $x=0$ crosses for the first time the energy level  
 $\mathcal{E}= {E}/{\sqrt{2 \Gamma}}$ at the turning point $x^*(\mathcal{E})$. }
\label{Fig:schema}
\end{center}
\end{figure}

We now estimate $n_E(0)$. First, we rewrite the Schr\"odinger equation~\eqref{eigenEq} as
\begin{equation}
\label{eigenEq2} 
 \frac{d^2  n_E(x)}{d x^2}  + \frac{\sqrt{2 \Gamma}}{D}
 \left( \eta(x) - \mathcal{E}\right)    n_E(x) = 0\, ,
 \qquad \mathcal{E}= \frac{E}{\sqrt{2 \Gamma}} \, .
\end{equation}
The Brownian landscape  $\eta(x)$ starts at $\eta(x=0)=0$ and it will remain smaller than $\mathcal{E}$ 
up to a first crossing-point $x^*(\mathcal{E}) > 0$ such that  $\eta(x^*) = \mathcal{E}$
(see Figure~\ref{Fig:schema}). We know the typical scaling  $x^*(\mathcal{E}) \sim \mathcal{E}^2$.
Because $\eta(x)$  is a non-monotonous function of $x$, there will be many subsequent
crossing points. (For the Brownian landscape, there will be infinitely many crossing points
immediately following the first crossing \cite{BM}.) We are interested in the behavior in the vicinity 
of the origin, see Eq.~\eqref{PopTotale}, so these further crossing points play a little role. 

In the long-time limit, the dominant contribution into the integral in \eqref{PopTotale} is provided by 
large values of `energy' $E$. (We shall confirm this assertion below.) Therefore we can analyze
Eq.~\eqref{eigenEq2} using the WKB method \cite{Migdal,BO78}. 
On the interval  $x < x^*$, the WKB solution reads 
\begin{equation}
\label{Interior}
n_E(x) \sim \frac{1}{\left(\mathcal{E} -\eta(x)\right)^{1/4}}  \,  
  {\exp} \left( {\frac{-(2 \Gamma)^{1/4} }{\sqrt{D}} 
 \int_x^{x^*(\mathcal{E})}    } 
dx' \,   \sqrt{\mathcal{E} -\eta(x') } \right)  
\end{equation}
We now define $x' = \mathcal{E}^2 \tau$ and use the scaling property 
$\eta(\mathcal{E} ^2 \tau) \equiv  \mathcal{E}  \eta(\tau)$. Then, from 
the expression~\eqref{Interior}, we obtain (keeping again only the dominant exponential factor)
\begin{equation}
  n_E(0) \sim \exp\!\left(  -\frac{ {E}^{5/2}}{2 \Gamma \sqrt{D}} \, \int_0^{\tau^*}
d\tau \,   \sqrt{B_\tau } \right) 
\label{WKBresult}
\end{equation}
where $\tau^*$ represents the first moment when the Brownian
motion $B_\tau = 1 -\eta(\tau)$ that starts at 1 at $\tau=0$, crosses the origin: 
$B_\tau>0$ for $\tau<\tau^*$ and  $B_{\tau^*}=0$.

We now  consider a typical  realization $\eta$ of the landscape.
The integral that appears in~\eqref{WKBresult} 
has a constant value $K_{\eta}$ that depends on the landscape. Thus
\begin{equation}
  n_{E,\text{typ}}(0) \sim  
 \exp\!\left(  - K_{\eta} \frac{ {E}^{5/2}}{2 \Gamma \sqrt{D}}  \right) \, .
\end{equation}
Substituting this expression in the spectral decomposition~\eqref{PopTotale}
leads to 
\begin{equation}
 n_{\text{typ}}  \sim   \int dE \,  \exp\!\left(Et 
 - K_\eta \frac{ {E}^{5/2}}{2 \Gamma \sqrt{D}}  \right) 
 \sim    \exp\!\left[\frac{3}{5} \left(\frac{4}{ 5 K_{\eta}}\right)^{2/3}
   ( \Gamma^2 D t^5)^{1/3} \right].  
\label{PopTypiqueTotale}
\end{equation}
The second asymptotic in \eqref{PopTypiqueTotale} has been derived via the saddle-point method. 
The maximum of the integrand occurs at $E\sim t^{2/3}$ which diverges as $t\to\infty$; 
this justifies the use of the WKB solution  \eqref{Interior}. The final result \eqref{PopTypiqueTotale} qualitatively agrees with the prediction of Eq.~\eqref{density} which was established using heuristic arguments. 

We now determine  the value of the average population $\langle n \rangle$, 
where the average is taken over all possible
realizations of the noise. First, we  need to  calculate
\begin{equation}
\langle  n_E(0)  \rangle    \sim  
\left\langle \exp\!\left(  -\frac{ {E}^{5/2}}{2 \Gamma \sqrt{D}} \, \int_0^{\tau^*}
d\tau \,   \sqrt{B_\tau} \right)  \right\rangle  \, ,
 \label{Brfunctional}
\end{equation}
where the expectation value is taken over all Brownian paths starting at 1 at $\tau=0$
and vanishing at $\tau^*$ for the first time.   This expression is the average of 
a first passage exponential  functional of the  Brownian motion and its value can be 
determined using the general method described e.g. in Ref.~\cite{Satya}. 
The procedure applies to a functional of the form 
\begin{equation}
\label{Q_def}
  Q(x_0) \equiv Q(x_0; p, \mathcal{U}) = \left\langle \exp\!\left(  -p \, \int_0^{\tau^*}
d\tau\,  {\mathcal  U}(B_\tau) \right)  \right\rangle  
\end{equation}
that involves an arbitrary smooth function ${\mathcal  U}(B_\tau)$ replacing $\sqrt{B_\tau}$ 
which appears in \eqref{Brfunctional}. The normalized Brownian process starts at any 
$x_0 >0$  and as a function of $x_0$, the functional \eqref{Q_def} satisfies the backward 
Fokker-Planck equation 
\begin{equation}
 \frac{1}{2}\frac{d^2  Q(x_0)}{d x_0^2} - p\, {\mathcal  U}(x_0)Q(x_0) =0
\label{BackFP}
\end{equation}
with the boundary conditions
\begin{equation}
 Q(x_0 =0) = 1 \,\, \hbox{ and }  \,\,  Q(x_0 \to +\infty) = 0 \, .
\label{bcFP}
\end{equation}
Our problem, equation~\eqref{Brfunctional}, perfectly fits into this framework. Thus we 
must solve the following differential equation:
\begin{equation}
  \frac{1}{2}\frac{d^2  Q(x_0)}{d x_0^2} - p \sqrt{x_0}Q(x_0) =0
  \,\,\hbox{ where } \,\,  p = \frac{ {E}^{5/2}}{2 \Gamma \sqrt{D}} \,.
\label{FPtosolve}
\end{equation}
We need  to calculate  $Q(1)$ because the Brownian paths in 
equation~\eqref{Brfunctional} start at $x_0 =1$. The asymptotic behavior of the solution 
of  Eq.~\eqref{FPtosolve} can again be obtained via the WKB method. 
Writing $Q = {\rm e}^S$ we find $S(x_0) = - \tfrac{4}{5}\,  \sqrt{2p}\,  (x_0)^{5/4}$ 
in the leading order. Therefore
\begin{equation}
\langle  n_E(0)  \rangle \sim  
\exp\!\left( - \frac{4 {E}^{5/4}  }{5 \,\Gamma^{1/2} D^{1/4} }  \right) \, .
\end{equation}
Inserting this result into equation~\eqref{PopTotale} and evaluating the integral
by saddle-point method we arrive at
\begin{equation}
\label{WKB_basic}
\langle  n   \rangle \sim 
\exp\!\left( \tfrac{1}{5} D \Gamma^2 t^5 \right) \, .
\end{equation}
This exact asymptotic qualitatively agree with Eq.~\eqref{av_cont} which was derived
using qualitative arguments. 

As a side remark, we notice that the growth laws in judiciously chosen deterministic landscapes 
$\eta(x)$ shed light on the growth laws in random landcapes. For instance, 
\begin{itemize}
\item  If we consider  $\eta(x) =  \kappa x$, then equation~\eqref{Diff-Lang}
admits an exact solution
\begin{equation*}
n(x,t) =  \frac{1}{\sqrt{4\pi D t}} 
  \,\, \exp\!\left(-\frac{(x - \kappa D t^2)^2 }{4 D t} + \frac{\kappa^2 D t^3}{3}  \right)
\end{equation*}
representing a Gaussian profile with uniformly accelerated center and with total mass growing as ${\rm e}^{t^3}$. This solution can be obtained by the previously described spectral analysis: $n_E(x) =  {\rm Ai}(E-x)$ where  ${\rm Ai}$  is the Airy function. This example can be used as a template
for  applying the WKB method. Note that this solution is well-known
in quantum mechanics as an `Airy Packet', that spreads without changing its form \cite{Berry}.

\item  For   $\eta(x) = \kappa x^2$, the path integral~\eqref{pathintegral}  is quadratic
and can be calculated exactly \cite{Schulman}.  The solution is given by 
 $$  n(x,t) =  \left( \frac{\kappa}{D} \right)^{1/4} \, 
  \frac{1}{ \sqrt{2\pi \sin(2\, (D\kappa)^{1/2}\, t)} }   
 \exp\left( -\frac{\sqrt{\kappa}\,  x^2}{2 \sqrt{D}}
 \frac{ \cos(2\, (D\kappa)^{1/2}\,t)}{ \sin(2\, (D\kappa)^{1/2}\,t)} \right)\, . $$
The total population corresponding to this solution, 
$N = 1/\sqrt{\cos[2\, (D\kappa)^{1/2}\,t]}$, blows up at  the finite
time $t_c = \, (D\kappa)^{-1/2}\, \frac{\pi}{4}\,. $

\item    For  $\eta(x) = \kappa\,{\rm sign}(x) |x|^\alpha$, with $0< \alpha < 2$,
the WKB analysis can be carried out again and one finds that the total population increases as
 \begin{equation*}
 N\sim   \exp\!\left( \frac{2-\alpha}{2+\alpha} 
 \left( A(\alpha)^{2\alpha}  D^{\alpha} \kappa^{2}
 t^{2+\alpha} \right)^{\frac{1}{2-\alpha}} \right)\, ,   \quad   
 A(\alpha) = \frac{ 2 \alpha^2 \Gamma(3/2 + 1/\alpha)}
  { (2 +\alpha) \Gamma(3/2)\Gamma( 1/\alpha)} \, .
\end{equation*}
\end{itemize}
For $\alpha = 1/2$, we arrive at the  $\exp(t^{5/3})$ growth  that was obtained 
for a typical  Brownian landscape.  For $\alpha = 1$, we recover  the 
$\exp\left(\kappa^2 D t^3/3  \right)$  growth of  the linear landscape.
For $0 < \alpha <1 $, we obtain the same behavior as  for
a  self-affine disordered landscape with Hurst exponent $H = \alpha$.

\subsection{Population Growth in High Dimensions}
\label{2d:corr}

When the dimension of the underlying substrate exceeds one, various types of random landscapes can arise and there is no single landscape which is as natural as the Brownian landscape in $1d$.  
Perhaps the closest analog of the one-dimensional Brownian landscape is a random landscape 
that arises by taking the `height' variable $\eta({\bf R})$ to be  a Gaussian free field.
This class of random manifolds has been widely studied (see e.g. \cite{SS} and references therein).
The Gaussian free field on a discrete lattice of dimension $d$
is defined  by assigning the Gaussian probability to the configuration $\eta({\bf R})$: 
\begin{equation}
\label{Gauss} 
\exp\left\{-\sum [\eta({\bf R}\pm {\bf e}_\alpha)-\eta({\bf R})]^2\right\} \, ,
\end{equation}
where the  sum  runs over ${\bf R}\in \mathbb{Z}^d$ and over  ${\bf e}_\alpha$, which are  
the unit vectors in the $d$ possible directions.  In the continuum limit,
the summation is replaced by integration, $\int (\nabla \eta)^2\,d{\bf R}$. 
Further, one has to introduce a small-scale cut-off to regularize  correlation functions on
short distances \cite{SS}. The Gaussian free field is 
characterized by  the following height-correlation function: 
\begin{eqnarray}
\label{height_2d}
 \langle [\eta({\bf R})-\eta({\bf 0})]^2\rangle &\sim&
\ln R \, \,\, \quad {\rm for  } \,  d=2 \,  ,  \\
\label{height_highd}
\langle [\eta({\bf R})-\eta({\bf 0})]^2\rangle &\sim& {\rm finite}
 \,   \quad {\rm for } \, d >2 \, .
\end{eqnarray}
One can also define  random landscapes by  using the
Edwards-Wilkinson growth process \cite{HHZ} which has
the same 2-point correlation functions as above.

\subsubsection{Typical  Growth}

We estimate the typical growth using the same argument as in one
dimension, namely by balancing the typical maximal value of the noise over
a radius $R$  with the probability for the particles to visit a droplet of size   $R$:
\begin{equation}
\label{typ_estimate_2d}  
n_{\rm typ}\sim \exp\!\left[t\sqrt{\ln R}
-\frac{R^2}{4t}\right] \, .
\end{equation} 
Maximizing with respect to  $R$ gives $R\sim t (\ln t)^{1/4}$ leading to 
\begin{equation}
\label{typ_2d}  
n_{\rm typ}\sim e^{t\sqrt{\ln t}} \, .
\end{equation}

In dimensions strictly higher than 2, if we consider a random
landscape generated by a Gaussian free field  then the landscape 
is statistically flat, i.e.,  its  width does not vary with the macroscopic length scale $R$
(it depends in fact on the microscopic cut-off). Therefore, the typical  population grows exponentially.

To summarize, we have the following growth laws for the typical population
under a random  landscape
\begin{equation} 
   n_{{\rm typ}} \sim
 \begin{cases}
e^{t^{5/3}} &\text{ for }  d =1 \\
e^{t\sqrt{\ln t}}    &\text{ for } d =2  \\ 
e^{t}  &\text{ for }  d \ge 3  \, .
\end{cases}
\label{HighDim}
\end{equation}

Finally, we note that random surfaces characterized by the logarithmically growing
height-correlation function \eqref{height_2d} are almost flat. One  can 
consider random surfaces with algebraic height-correlation
function  (see e.g. \cite{K}) given by  $ \langle [\eta({\bf R})-\eta({\bf
0})]^2\rangle \sim R^\zeta .$   We then  obtain
$ n_{\rm typ}\sim \exp\!\left[t R^{\zeta/2}-\frac{R^2}{4t}\right]$, from which we conclude that 
\begin{equation} 
n_{\rm typ}\sim
\exp\!\left[t^{(4+\zeta)/(4-\zeta)}\right] \, .
\end{equation}

\subsubsection{Average Growth}

The typical growth laws obtained above correspond  to  a given realization of 
the Gaussian free field  $\eta({\bf R})$. The average growth of the population is again 
dominated by the very rare fluctuations of the landscape.
To derive the average growth we need,  as in one dimension,  to keep two free parameters:
the size $R$ of the droplet visited by the particles and the maximum value  $M$
reached by   $\eta({\bf R})$ over this droplet.

In two dimensions, a heuristic  estimate is found by  taking a Gaussian distribution for $M$
of variance $\ln R$: 
\begin{equation}
\label{av_2dim}  \langle n\rangle\sim \exp\!\left[t M - \frac{M^2}{2\,\ln R}
-\frac{R^2}{4t}\right] \, . 
\end{equation}
Maximizing  with respect to  $M$ and $R$,  we find 
\begin{equation}
 \langle n\rangle\sim e^{t^2\, \ln t} \, .
\label{av_2d} 
\end{equation}
A rigorous  derivation of this estimate is perhaps a challenging problem. Nevertheless, 
we can justify  why the distribution of the maximum  value $M$ of a Gaussian free field  has a
Gaussian tail. Let us first revisit the  $d=1$ case. The full 
distribution of the maximum of a Brownian process is of course
well-known \cite{BM}, yet we want to deduce its tail in a simple way that will admit a generalization
to the Gaussian free field. The tail of the maximum distribution can be retrieved  by the following
simple reasoning: let us consider a Brownian path $h(x)$  of length
$L$ with $h(0)=0$  and let  $M$  be its maximum value; 
when $M$ is large, the  maximum must be  reached in the vicinity of
the end of the path and  such a path contributes by a weight of
$\exp(-\tfrac{1}{2}\int (h'(x))^2 dx)$. The optimal path that has the largest  weight is
therefore obtained by minimazing the integral $\int_0^L (h'(x))^2 dx$. This gives $h''=0$, which in conjunction with $h(0) = 0$ and $h(L) = M$, leads to $h(x)= Mx/L$.
Substituting this expression in the exponential weight, we arrive at the tail $\exp( -M^2/2L)$ 
which we used in Eq.~\eqref{av_est}.

Turning to two dimensions, let us consider a Gaussian random surface $h(x,y)$ over
a circular disk of radius  $R$. As in $d=1$ we expect  the 
maximum value $M$ of the height $h$  to be reached on the rim of the disk.
Supposing  that the optimal surface is rotationally invariant,
its  statistical  weight is given by $\exp(-\tfrac{1}{2}\int (\frac{\partial h}{\partial r})^2 r  dr)$
in radial coordinate $r$. Optimizing this weight with the
constraints $h(0) = 0$ and $h(R) = M$ leads to the Euler-Lagrange
equation:
\begin{equation*}
 \frac{d}{dr} \left( r \frac{dh}{dr} \right) = 0 \,.
\end{equation*}
The solution to this equation is $h(r) = M \frac{\ln r}{\ln R}$ and 
the weight of this optimal path is given by
\begin{equation*}
\exp\left(- \frac{M^2}{2(\ln R)^2} \int_1^R \frac{ dr}{r}\right) =  
             \exp\left( -\frac{M^2}{2\ln R}\right) \, ,
\end{equation*}
where a short-length scale cut-off (we set it to unity) allows to avoid the small $r$ divergence.
This justifies the expression of the distribution of $M$ used in equation~\eqref{av_2dim}.

\begin{table}  \centering
  \begin{tabular}{|c| c | c |}
 \hline
 Growth in various  $d$ &  Typical &   Average \\
 \hline                      &         &     \\ 
One dimension        &   $e^{t^{5/3}}$   &   $e^{t^5}$   \\
                                &         &     \\   
 \hline                      &         &     \\ 
Two dimensions      &    $e^{t\sqrt{\ln t}}$   &     $e^{t^2 \ln t}$       \\
                               &         &     \\
 \hline                     &         &     \\
 Higher dimensions ($d>2$)   &        $e^{t}$   &          $e^{t^2}$  \\
                     &         &     \\
 \hline           
  \end{tabular}
\caption{Asymptotic growth laws in a Gaussian random landscape. 
In order to regularize short scale fluctuations in $d\geq 2$, the Gaussian free field is defined on a lattice. For $d=1$, the results were already  given in Table~\ref{tablesynthese}.}
\label{tableHighDim}
\end{table}

A similar reasoning can be carried out in higher dimensions.
In order to estimate  the average growth we again need to know the distribution 
of the maximum. Proceeding as above we find that in three dimensions,
the spherically symmetric optimal landscape satisfies  
$\frac{d}{dr}\left( r^2 \frac{dh}{dr}  \right) = 0$  with
respect to the radial coordinate $r$.  This leads to 
$h(r) = M \frac{R(r-1)}{r(R-1)}\to M\big(1-r^{-1}\big)$ (taking again
the microscopic cut-off to be unity). The corresponding weight 
behaves as $\exp(-M^2/2)$; there is no dependence 
on $R$  because the interface is flat at large scales. Using this expression
for the maximum distribution, we obtain the average growth of the population
 $$  \langle n\rangle\sim \exp\!\left[t M - \frac{M^2}{2}
-\frac{R^2}{4t}\right] \, . $$
in three dimensions. Maximizing with respect to $M$   we arrive at 
\begin{equation}
\label{av_3dim}
\langle n\rangle \sim e^{t^2} \, .
\end{equation}
Note that we do not need to  optimize with respect to $R$ because the distribution of $M$, for large values of $M$, is  independent of  $R$. A similar reasoning can be carried out in higher dimensions
and we find that the  behavior is also  given  by \eqref{av_3dim} for all $d>2$.

Table~\ref{tableHighDim} summarizes our results for the typical and the average  population
growth in the situation when the random landscape is described by a Gaussian free field.

\section{Conclusion}

In this work, we have investigated the evolution of a population of non-interacting particles 
that undergo diffusion and birth/death. The latter depends on the environment, for example,  the 
distribution of nutrients that  defines a landscape which is assumed to be stationary.  
Different  statistical properties of this landscape lead to a number of laws for 
the growth of the population.  In most of the cases, the total  population increases 
in a faster-than-exponential manner. This behavior is due to  two features of the noise, 
namely its multiplicative nature and the lack of upper bound. Another striking feature 
is the huge difference between typical and average behaviors.  In order to determine 
the average  growth law, one has to consider all possible realizations of the random 
landscape and  the  average  is dominated by very rare configurations. 
Thus extremal statistics play an important role.  Some of our analysis relies 
on heuristic arguments. In one dimension  we have performed asymptotically 
exact calculations in the situation where  the noise is described by a Brownian 
process. We have shown that the  determination of  the average population growth 
reduces to calculating a first passage exponential functional of the Brownian motion. 
This problem can be solved by using  the Backward Fokker-Planck equation. 
The asymptotically exact results agree with heuristic predictions. 

Although the basic stochastic differential equation \eqref{Diff-Lang}  has been studied 
for almost thirty years, there are still many open problems. Even in one dimension, 
it would be interesting to generalize the quantitative  approach  applicable 
to the Brownian landscape to other random landscapes
such as those generated by a fractional Brownian motion. In higher dimensions 
($d \ge 2$),  little is rigorously and/or exactly known. To appreciate the challenge, 
one can think of  the  somewhat related problem of the localization of a quantum particle.  

On a more practical side, it could be interesting  to study the transient regime when the system 
is still far from the final asymptotic stage. In the long time limit, non-linear saturation effects, 
that are ignored in our model,  can start playing a prominent role. 
Adding nonlinear terms to the basic equation~\eqref{Diff-Lang} will totally modify 
its properties in the asymptotic regime. This is a challenging
mathematical problem that deserves further analysis.

\bigskip
The work of PLK has been supported by NSF Grant No.\ CCF-0829541.
We are thankful to M. Bauer, F. David, B. Duplantier,  S. Mallick, and S. Redner for suggestions, help, 
and encouragement.

\end{document}